%% file: main.tex
\documentclass[10pt]{IEEEtran}
\usepackage[nocompress]{cite}
\usepackage{amsmath,amssymb,amsfonts}
\usepackage{graphicx}
\usepackage{textcomp}
\usepackage{xcolor}
\usepackage[T1]{fontenc}
\usepackage{booktabs,tabularx,array}
\usepackage[hidelinks]{hyperref}
\graphicspath{{figures/}}
\newcolumntype{Y}{>{\raggedright\arraybackslash}X}

\begin{document}
\title{Atlas: Accelerating Quantum FTQC Compilation with Optimized Algebraic Rotation Synthesis}
\author{Anbang Wu, Pu Pang, and Wei Geng%
\IEEEcompsocitemizethanks{%
\IEEEcompsocthanksitem Anbang Wu and Pu Pang are with Shanghai Jiao Tong University
(e-mail: \mbox{anbang@cs.sjtu.edu.cn}; \mbox{pangpu@cs.sjtu.edu.cn}).
\IEEEcompsocthanksitem Wei Geng is with South China University of Technology
(e-mail: \mbox{geng05wei25@163.com}).}}
\IEEEtitleabstractindextext{\input{sections/abstract}}
\maketitle
\IEEEdisplaynontitleabstractindextext
\IEEEpeerreviewmaketitle
\input{sections/introduction}
\input{sections/background}
\input{sections/design}
\input{sections/implementation}
\input{sections/evaluation}

\input{sections/discussion}

\input{sections/related-work}
\bibliographystyle{IEEEtran}
\bibliography{references}
\end{document}

%% file: sections/abstract.tex
\begin{abstract}
As fault-tolerant quantum computing emerges, Clifford+T gate basis becomes a popular choice where we need to decompose single-qubit rotation gates into gates in this gate basis. Quantum rotation synthesis may lead to remarkable overhead in quantum circuit compilation. Considering variational quantum eigensolver (VQE) applications as an example, a conventional UCCSD ansatz in quantum chemistry can generate $O(n^4)$ rotations for $n$ qubits. Each optimization iteration updates the rotation angles, making rotation synthesis a recurring cost that can delay the VQE training loop. We identify two sources of delay in rotation synthesis: redundant computation during synthesis and uneven synthesis times that leave workers waiting for slow requests. Atlas addresses these costs with three optimizations: search pruning skips invalid synthesis candidates through statistical inference; adaptive arithmetic avoid unnecessary high-overhead big integer/float arithmetic when decomposing the target rotation gate; batch scheduling fills the gap between the synthesis time of varied rotations. Across various benchmarks, synthesis method, and compilation machines, Atlas can consistently accelerate the rotation synthesis process by up to 12.65x.
\end{abstract}

%% file: sections/introduction.tex
\section{Introduction}
\label{sec:introduction}

As quantum hardware and computational workloads scale, support for fault-tolerant quantum computing (FTQC) becomes increasingly important in quantum compilation. The Clifford+T gate set is a common target for this setting, providing a discrete universal basis for implementing quantum algorithms. However, many applications are typically expressed using arbitrary-angle rotations that generally require approximation in this basis. The compiler must therefore translate these rotations into Clifford+T gate sequences under a specified error tolerance. This process, known as rotation synthesis, connects application-level quantum circuits to their fault-tolerant implementations.

Rotation synthesis introduces two coupled sources of compilation cost: the number of rotations and the accuracy required for each rotation. For a single rotation, algebraic rotation synthesis methods \cite{ross2016optimal,bocharov2015pqf} like GridSynth \cite{ross2016optimal} has an expected classical runtime of $O(\operatorname{polylog}(1/\varepsilon))$ under a number-theoretic hypothesis, where $\varepsilon$ is the approximation tolerance \cite{ross2016optimal}. Larger circuits has more rotation gates and require more synthesis requests at tighter tolerances, contributing to huge synthesis overhead and compilation time. 

This cost is particularly significant for applications like the variational quantum eigensolver (VQE) \cite{peruzzo2014variational,mcclean2016theory}, adopted in various chemistry and physics simulation tasks. On the one hand, the number of rotation gates grow quickly in VQE applications. In half-filled unitary coupled cluster singles and doubles (UCCSD) workloads \cite{romero2019ucc}, rotation gates grow as $\Theta(n^4)$, with the qubit count $n$. The weight-four electronic qubit coupled cluster (QCC) \cite{ryabinkin2018qcc} and dense Sachdev--Ye--Kitaev (SYK-4) blocks \cite{maldacena2016syk} also exhibit quartic growth; increasing QCC from 10 to 20 qubits raises its rotation count from 8,400 to 193,800. On the other hand, VQE further updates rotation parameters throughout optimization \cite{mcclean2016theory}. With a fixed ansatz and mapping, compilation of the parameter-independent Clifford structure can be reused, but changed angles generate new synthesis requests. Because these angles depend on the optimization trajectory, synthesis cannot generally be completed offline and can increase the total training time.

These observations demands efficient rotation synthesis in the quantum compiler pipeline.
Unfortunately, by profiling on existing algebraic synthesizers, including GridSynth and PQF \cite{ross2016optimal,bocharov2015pqf}, we observe three recurring sources of performance bottlenecks despite their different search procedures:

\textbf{1) Unnecessary search exploration.} Existing algebraic rotation synthesizers enumerate algebraic candidates in successive layers of increasing T gate count in the gate sequence before decomposing an rotation gate into Clifford+T gates. Starting at the lowest layer and advancing sequentially spends time on layers that contain no admissible candidate. The relationships between layers provide opportunities to rule out this work, but a conventional traversal does not exploit them to skip entire empty-solution ranges. Repeating search preparation and candidate enumeration adds further overhead.

\textbf{2) Expensive arithmetic.} Tight approximation tolerances require accurate candidate calculations and exact algebraic computations. Existing implementations rely heavily on general-purpose multi-precision routines, even when the current operands fit in machine words or the requested tolerance permits a lower working precision. Repeated constant evaluation, arithmetic setup, and numerical-output conversion further increase the cost. Consequently, the implementation can spend substantial time on arithmetic whose generality or precision exceeds the needs of the current computation.

\textbf{3) Long-tail search latency.} Different rotations require different numbers of candidate tests and solving attempts. A small number of difficult searches can therefore delay completion of an entire batch even when rotations are synthesized simultaneously on parallel workers. Distributing independent rotations among workers does not remove the redundant work within each search, and fixed assignments leave some workers idle while others finish. Under a limited CPU budget, reducing per-request work and balancing the remaining computation are both necessary.

In this paper, we propose Atlas to accelerate algebraic rotation synthesis. Our key insight is to exploit structural patterns within and across synthesis searches to eliminate unnecessary computation and mitigate long-tail search latency. The reusable structure spans candidate layers, arithmetic operations, and compatible steps in independent searches. Atlas combines three complementary optimizations.

\textbf{1) Search pruning.} We exploit statistical profiling results to skip layers established to be empty and reject invalid candidates before expensive candidate solving. Initial probes test layers for admissible candidates and retain enumeration state for reuse in the subsequent search. This optimization significantly shortens the overall search process.

\textbf{2) Adaptive arithmetic.} We select arithmetic routines according to operand width and working precision. Machine-word operations handle supported small operands, while multi-precision routines handle larger values. Constants are shared at the required precision, and specialized numerical-output routines reduce conversion costs. These changes reduce arithmetic overhead while retaining reliable checks against the requested synthesis tolerance.

\textbf{3) Batching and scheduling.} We divide the overall rotation synthesis process into modular units and batch these smaller-scale operations across rotations to fill in idle slots on parallel CPU workers. Further, combining shared computation with dynamic scheduling reduces repeated work and limits the idle time caused by uneven search lengths.

We implement Atlas in GridSynth and the PQF synthesizer, and evaluate it on random rotations and five VQE workload families across two CPU platforms. At matched worker counts, Atlas achieves aggregate VQE synthesis speedups of $5.57$--$5.88\times$ for RustGridSynth and $5.66$--$12.65\times$ for PQF. Consider the full compiler pipeline which consists of optimizations more than rotation synthesis, Atlas yields end-to-end compilation speedups of $1.97$--$2.37\times$ on VQE workloads at a per-rotation tolerance of $10^{-8}$, over Gridsynth; and $1.97$--$2.37\times$ end-to-end compilation speedups over PQF.

The main contributions of this paper are as follows:
\begin{itemize}
\item We characterize the recurring compilation cost of rotation synthesis in parameterized quantum workloads and identify redundant search, expensive arithmetic, and uneven search latency as optimization opportunities.
\item We develop search pruning, adaptive arithmetic, and batching with dynamic scheduling to exploit structural patterns within and across synthesis requests.
\item We implement these techniques for deterministic and probabilistic synthesis, and evaluate their performance, interactions, and circuit-cost tradeoffs through performance comparisons and ablation studies on two CPU platforms.
\end{itemize}

%% file: sections/background.tex
\section{Background and Motivation}
\label{sec:background}

\subsection{Rotation Synthesis}
Quantum programs commonly express single-qubit operations as parameterized rotations. Fault-tolerant compilation replaces these operations with gates from a discrete basis. We consider Clifford+T, where the single-qubit Clifford gates are generated by H and S, and $T=\operatorname{diag}(1,e^{i\pi/4})$. Clifford basis changes reduce $X$ and $Y$ rotations to $Z$ rotations. The synthesis task is to construct a gate sequence $C$ satisfying
\begin{equation}
R_Z(\theta)=\operatorname{diag}(e^{-i\theta/2},e^{i\theta/2}),\quad
\|C-R_Z(\theta)\|\leq\varepsilon,
\label{eq:rotation-goal}
\end{equation}
where the norm is the operator norm and $\varepsilon$ is the synthesis tolerance. The number of T gates measures circuit cost, while compilation time measures the classical work needed to find the circuit. 

\begin{figure*}[t]
\centering
\includegraphics[width=\textwidth]{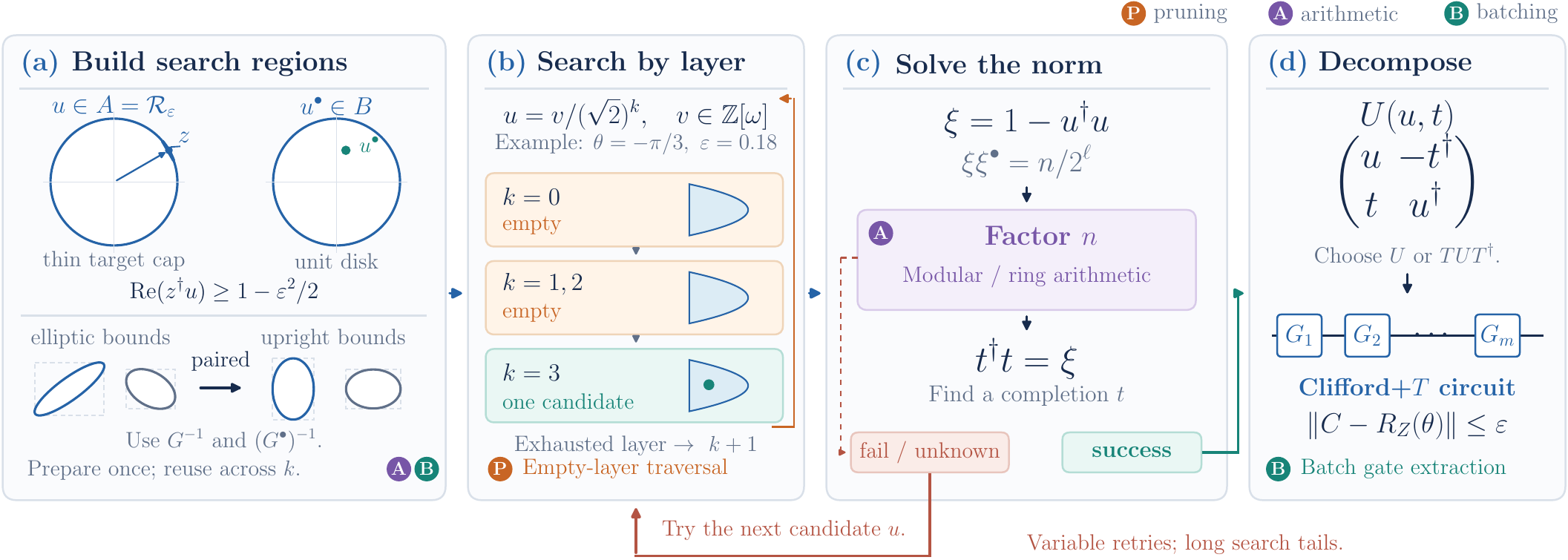}
\caption{GridSynth's fixed-phase search and its computational bottlenecks~\cite{ross2016optimal}. (a) Paired regions and enclosing ellipses prepare bounded enumeration. (b) The example $\theta=-\pi/3$, $\varepsilon=0.18$ has no geometric candidates for $k<3$ and one at $k=3$; cap insets stretch the radial direction for visibility. (c) Failure or an unresolved attempt retries the next candidate; an exhausted layer advances $k$. (d) Exact decomposition follows successful norm completion. Enclosure and circuit icons are schematic; P/A/B mark Atlas optimization opportunities.}
\label{fig:gridsynth-search}
\end{figure*}

We use GridSynth \cite{ross2016optimal} as an representive to illustrate the algebraic rotation synthesis process.  Figure~\ref{fig:gridsynth-search} illustrates its fixed-phase formulation. With $\omega=e^{i\pi/4}$, the search constructs
\begin{equation}
U=\begin{pmatrix}u -t^{\dagger}\\t u^{\dagger}\end{pmatrix},
u,t\in\mathbb D[\omega]=\mathbb Z[1/\sqrt{2},i],
\label{eq:algebraic-unitary}
\end{equation}
These algebraic entries permit exact Clifford+$T$ decomposition once $U$ is unitary~\cite{kliuchnikov2013exact}. Finding $U$ therefore requires both geometric approximation and algebraic completion.

\textbf{Geometric preparation.} Let $z=e^{-i\theta/2}$. Figure~\ref{fig:gridsynth-search}(a) shows the target cap $A$ and a second unit disk $B$, which demonstrates
constraints necessary for unitarity and makes this bounded search finite. GridSynth encloses the regions in ellipses and computes a grid-preserving coordinate transformation that improves their alignment with bounding rectangles. Enumeration uses the bounding rectangles and checks points against the actual regions.

\textbf{Layer and candidate enumeration.} The search visits candidates in increasing least denominator exponent, starting at $k=0$ (Fig.~\ref{fig:gridsynth-search}(b)). Here, $\mathcal C_k$ contains all candidates with denominator exponent at most $k$ (thus about $k$ T gates); the new layer is $\mathcal C_k\setminus\mathcal C_{k-1}$, with $\mathcal C_{-1}=\varnothing$. This distinction matters when reasoning about empty layers. A layer without an accepted result advances the search to the next exponent.

\textbf{Norm completion.} Each geometric candidate $u$ must admit an algebraic entry $t$ satisfying
\begin{equation}
t^\dagger t=\xi=1-u^\dagger u,\qquad t\in\mathbb D[\omega].
\label{eq:norm-equation}
\end{equation}
The solver uses integer factorization to solve the norm equation (Fig.~\ref{fig:gridsynth-search}(c)). A candidate can fail this test; a practical implementation can also abandon a difficult factorization. Both cases return to candidate enumeration. Exhausting a layer advances $k$. The number and duration of these attempts vary across rotations.

\textbf{Exact decomposition.} After finding $(u,t)$, the algorithm selects $U$ or $TUT^\dagger$ for the lower $T$-count and exactly decomposes the selected rotation gate by \cite{kliuchnikov2013exact} (Fig.~\ref{fig:gridsynth-search}(d)). 

\subsection{Computational Structure and Opportunities}
\begin{figure*}[t]
\centering
\includegraphics[width=\textwidth]{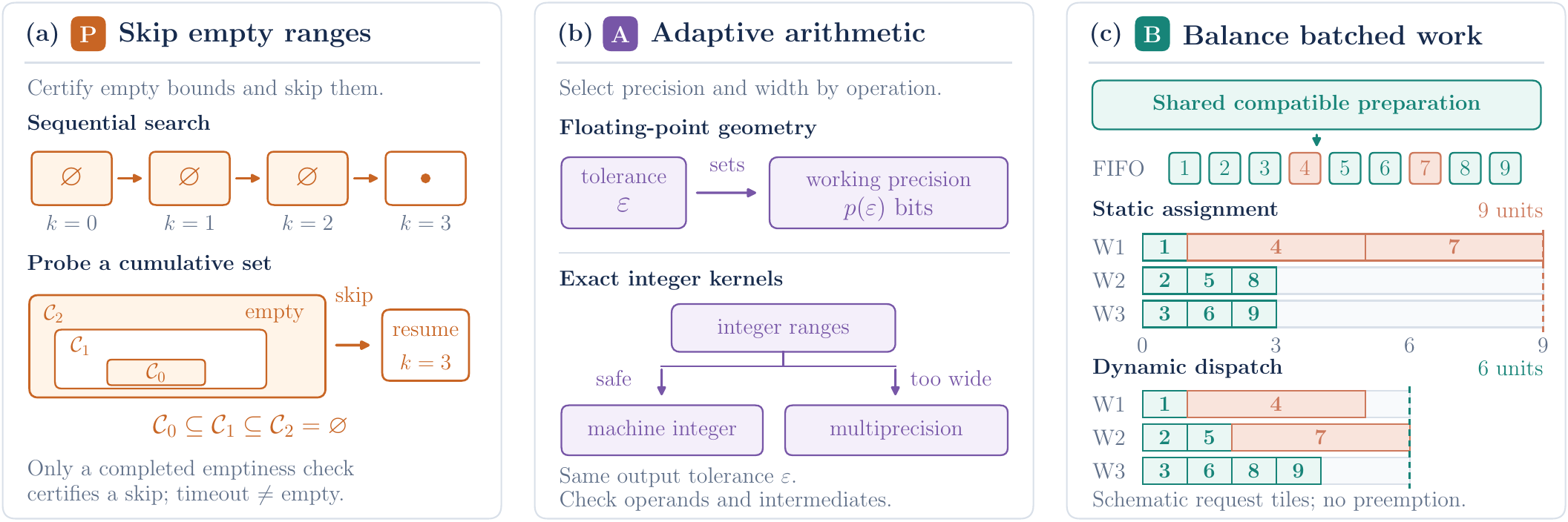}
\caption{Optimization opportunities exposed by the synthesis process. (a) P: probe cumulative candidate sets, skip certified empty ranges, and reuse enumeration state. (b) A: select arithmetic according to numerical precision and exact operand width. (c) B: batch compatible operations and dynamically dispatch small request batches to workers. The set diagrams, arithmetic icons, and execution timelines are schematic illustrations, not measured profiler data.}
\label{fig:gridsynth-opportunities}
\end{figure*}

Figure~\ref{fig:gridsynth-opportunities} connects these operations to optimization opportunities and reveals insights of Atlas. 

\textbf{Search pruning (P).} Since $\mathcal C_k\subseteq\mathcal C_{k+1}$, establishing that $\mathcal C_k$ is empty rules out all smaller denominator bounds. This supports skipping  empty ranges and retaining probe results.

\textbf{Adaptive arithmetic (A).} The search process undergoes with varying operand widths. Shared constants and specialized arithmetic can reduce overhead when their precision and range requirements are satisfied.

\textbf{Batching and scheduling (B).} Independent rotations repeat compatible operations but follow different retry paths. Dynamic assignment of small request batches reduces idle time from uneven search lengths. 

\section{Related Work}
\label{sec:related}

Single-qubit synthesis reduces T-cost through algebraic approximation and probabilistic constructions. GridSynth combines geometric search with exact Clifford+T decomposition~\cite{kliuchnikov2013exact,ross2016optimal}. RUS uses repeated probabilistic attempts, and PQF adds a deterministic fallback~\cite{paetznick2014rus,bocharov2015pqf}. Mixtures and magnitude approximation reduce gate cost~\cite{campbell2017mixing,kliuchnikov2023shorter}, with recent advances in general-unitary and small-angle synthesis~\cite{morisaki2025optimal,bothe2026smallangle}. Circuit-level methods merge rotations and coordinate their approximation~\cite{nam2018optimization,kishony2026optimal}.

Classical multi-qubit circuit numerical synthesis benefits from organized search, reuse, and specialized arithmetic~\cite{amy2013mitm,dimatteo2016parallelizing,paradis2024synthetiq,glaudell2026highperformance}. Atlas instead focuses on accelerating single-qubit algebraic rotation synthesis. CUDA-Q caches repeated rotations~\cite{cudaqRotationSynthesis} and is orthogonal to our work and can be combine with our acceleration techniques.

%% file: sections/design.tex
\section{Atlas Design}

Atlas accelerates rotation synthesis through search pruning (P), adaptive arithmetic (A), and batching with dynamic scheduling (B). It reduces per-request computation and long-tail latencies across independent rotations.

\subsection{Search Pruning and Probe Reuse}

\textbf{Key insights.} Candidate sets support prefix skipping, while saved iterators preserve useful probe work. For a fixed phase branch, the cumulative geometric sets satisfy $\mathcal C_k\subseteq\mathcal C_{k+1}$, where $k$ is the denominator bound. Exhausting the native enumeration of $\mathcal C_k$ advances the empty-prefix bound $k$.

Bounded probes focus exploration near occupied denominator bounds. Training statistics or analytic estimates select probe locations. An empty probe advances the lower bound; a discovered point supplies an upper bound. Budget exhaustion leaves both unchanged. A successful probe saves its first candidate and remaining iterator, so synthesis resumes in the original order and reuses completed geometric enumeration.

Starting rules and coefficient filters reduce work before norm solving. Atlas resumes beyond the exhausted prefix and lowers an empirical start hint when an earlier point is witnessed. Interval constraints on algebraic components and their conjugates reject infeasible coefficient pairs before full candidate construction.

\subsection{Adaptive Arithmetic and Output Construction}

\textbf{Key insights.} Integer width and numerical precision determine arithmetic cost, while repeated operations share preparation. Atlas selects kernels by operand range and reuses constants and conversion factors across compatible computations.

Width-specific kernels accelerate integer factorization's modular products. Moduli of at most 64 bits use 128-bit products; supported wider odd moduli use two-limb Montgomery arithmetic~\cite{montgomery1985modular}, with modulus-dependent setup shared throughout each walk. Multiprecision routines handle larger operands. Brent-style batching~\cite{brent1980improved} combines iterate differences into a modular product before each gcd, reducing gcd calls.

Precision-aware evaluation reduces numerical work. Atlas caches constants by precision and reuses region data at matching tolerance, scale, precision, and numerical policy. Paired sine and cosine evaluation shares argument reduction, while a tail bound stops additions that leave the stored result unchanged. Profiles shows that we can reduce working precision at tolerances $10^{-4}$, $10^{-5}$, and $10^{-6}$. In full PQF, this applies to correction synthesis; verification at the original precision triggers a full-precision retry if the branch error exceeds the target.

Direct decimal construction accelerates PQF's numerical outputs. For a binary value $m2^{-k}$ with integer $m$ and $k\geq0$, Atlas constructs $m5^k$, assigns scale $10^{-k}$, and rounds once to the interface's precision. Cached powers $5^{2^j}$ supply the multiplier; library conversion handles exponents outside the specialized range.

\subsection{Batching and Dynamic Scheduling}

\textbf{Key insights.} Rotations share local operations but have uneven search lengths. Atlas groups compatible operations within request tiles to share setup, then dynamically assigns tiles to balance work across CPU workers.

Taking the GridSynth as an example,
geometry batching shares transformations while preserving each request's state. Each reduction wave independently selects a transformation for every request, then groups requests by integer parameter under matching precision and policy. For the ellipse $ax^2+2bxy+dy^2$, the shear $x=x'+qy'$, $y=y'$ gives
\begin{equation}
a'=a,\qquad b'=aq+b,\qquad d'=aq^2+2bq+d.
\label{eq:atlas-shear}
\end{equation}
Requests share the prepared products, update their own coefficients and exact coordinate maps, and regroup in the next wave.

Compact integer states enable batched gate construction. Atlas stores accepted gate matrices as eight integer coefficients plus denominator exponent and phase. Low-order residue bits select gate-emitting reduction actions. Deterministic tiles batch geometry and decomposition around independent candidate searches. Full PQF groups primary matrices by precision and correction requests by precision and correction tolerance.

Dynamic scheduling distributes pending tiles as workers finish. Workers take tiles from a shared queue and return indexed results for input-order reconstruction. Smaller tiles improve load balance; larger tiles amortize dispatch and grouping. Our profiling selects the tile size for the available CPU budget.

%% file: sections/evaluation.tex
\providecommand{\evaluationfigures}{figures}
\section{Evaluation}\label{sec:evaluation}

We evaluate Atlas's synthesis performance, the contribution of its optimizations, and their effect on circuit cost and compilation time. The experiments address four research questions:
\begin{itemize}
\item \textbf{RQ1:} How much does Atlas accelerate synthesis across tolerances, CPU platforms, and VQE workloads?
\item \textbf{RQ2:} Which optimizations contribute most to the speedup, and how do they interact?
\item \textbf{RQ3:} How does Atlas affect the output circuit's $T$-cost and the synthesizer's memory use?
\item \textbf{RQ4:} How does faster synthesis change estimated compilation time and the remaining bottleneck?
\end{itemize}

\subsection{Experimental Setup}

\textbf{Baseline and Platform.} We compare Atlas with the original implementation of each synthesis method under the same CPU and worker budget. The methods are IBM RustGridSynth~0.2.2 \cite{rsgridsynth2026} and a Rust implementation of one-round probabilistic synthesis with fallback (PQF) \cite{bocharov2015pqf}. Experiments run on AMD EPYC~7742 and Intel Xeon Platinum~8562Y+ (128G RAM) hosts with 4, 8, or 16 parallel workers.

\textbf{Benchmarks.} The workloads cover both a broad precision range and complete rotation lists from VQE circuits. Random tests combine twelve tolerances, $\varepsilon=10^{-4},\ldots,10^{-15}$, with batches of 2,048, 4,096, or 8,192 rotations and all three worker counts. VQE tests cover five circuit families at 10 and 20 qubits (Table~\ref{tab:eval-workloads}), tolerances of $10^{-4}$ and $10^{-8}$ (commonly-adopted in existing works \cite{ross2016optimal,campbell2017mixing,kliuchnikov2023shorter}), and the same worker counts and rotation batch sizes. Their angles are sampled independently from approximately $[-2\pi,2\pi)$.

\begin{table}[t]
\centering
\caption{VQE workload information. QCC \cite{ryabinkin2018qcc} denotes qubit coupled cluster; UCCSD \cite{romero2019ucc} denotes unitary coupled cluster with single and double excitations; SYK-4 is the four-body Sachdev--Ye--Kitaev model \cite{maldacena2016syk}.}
\label{tab:eval-workloads}
\begin{tabular}{lrr}
\hline
Workload & 10 qubits & 20 qubits\\
\hline
Four-spin glass & 210 & 4,845\\
Anharmonic vibrations & 210 & 4,845\\
Electronic QCC & 8,400 & 193,800\\
Half-filled UCCSD & 850 & 16,400\\
Dense SYK-4 & 4,845 & 91,390\\
\hline
\end{tabular}
\end{table}

\textbf{Metrics.} Synthesis time measures request dispatch, circuit synthesis, and result collection. We average timing repeats within each case and sum these means over the cases being compared. Aggregate speedup is the original total divided by the Atlas total.

\subsection{RQ1: Synthesis Performance}

Atlas accelerates all 1,152 tested configurations, with aggregate speedups of $5.57$--$13.34\times$. Table~\ref{tab:eval-speedup} summarizes the results for each CPU and workload class. Random rotations gain $6.24$--$9.53\times$ with RustGridSynth and $10.52$--$13.34\times$ with PQF. For VQE, RustGridSynth achieves $5.57$--$5.88\times$, while PQF achieves $5.66$--$12.65\times$.

\begin{table}[t]
\centering
\caption{Aggregate synthesis speedup over each method's original implementation. Ratios divide summed per-case mean times.}
\label{tab:eval-speedup}
\begin{tabular}{llrr}
\hline
CPU & Workload & RustGridSynth & PQF\\
\hline
EPYC & Random & $9.53\times$ & $13.34\times$\\
Xeon & Random & $6.24\times$ & $10.52\times$\\
EPYC & VQE & $5.88\times$ & $12.65\times$\\
Xeon & VQE & $5.57\times$ & $5.66\times$\\
\hline
\end{tabular}
\end{table}

\begin{figure}[t]
\centering
\includegraphics[width=\columnwidth]{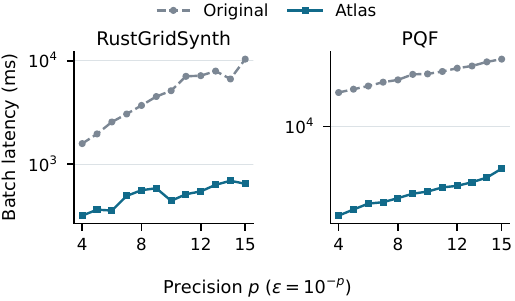}
\caption{Random-rotation synthesis time with eight workers and 4,096 requests. Each point averages three runs; $p$ denotes tolerance $10^{-p}$.}
\label{fig:eval-runtime}
\end{figure}

The performance gain persists as the requested tolerance tightens. Figure~\ref{fig:eval-runtime} fixes the workload at 4,096 random rotations and eight workers on both CPU platforms, exposing the effect of precision. Summing the mean batch times across all twelve tolerances, Atlas reduces RustGridSynth time from 61.81 to 6.14 seconds and PQF time from 386.74 to 28.35 seconds. Both methods improve at every tolerance, covering searches with different candidate counts and arithmetic precision requirements.

\begin{figure}[t]
\centering
\includegraphics[width=\columnwidth]{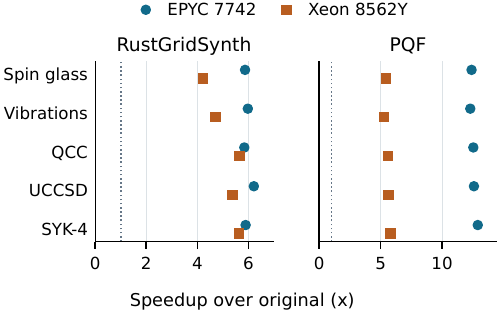}
\caption{Synthesis speedup for each VQE family. Each point aggregates 36 cases: two circuit sizes, two tolerances, three worker counts, and three input batch sizes on one CPU.}
\label{fig:eval-vqe}
\end{figure}

Atlas also speeds up every VQE family across a wide range of rotation counts. Figure~\ref{fig:eval-vqe} combines both circuit sizes, tolerances, worker counts, and batch sizes within each family. Family-level speedups range from $4.22$ to $6.21\times$ for RustGridSynth and from $5.28$ to $12.91\times$ for PQF across the two CPUs. The 10-qubit spin-glass and vibration circuits fit in one input batch, while the 20-qubit QCC circuit contains 193,800 rotations and spans many batches. Improvements across these cases show that the benefit extends from short request lists to sustained synthesis workloads.

\subsection{RQ2: Contributions of the Optimizations}

\begin{figure}[t]
\centering
\includegraphics[width=\columnwidth]{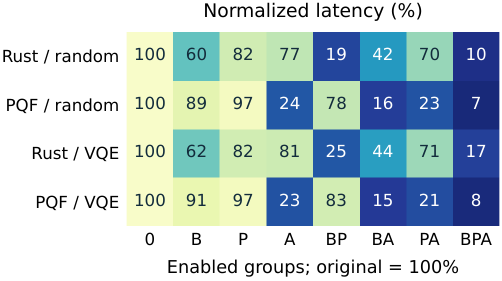}
\caption{CPU runtime as a percentage of the original implementation for all optimization combinations. B: batching and scheduling; P: search pruning; A: adaptive arithmetic. The label 0 denotes the original configuration.}
\label{fig:eval-ablation}
\end{figure}

Batching and pruning account for most of the time saved in RustGridSynth. Figure~\ref{fig:eval-ablation} compares all eight group combinations on both CPU platforms, normalized to the original runtime. Adding groups in the tuning order, B then P then A, attributes 45.2\% of the time saved on random workloads to B and 45.8\% to P; together they account for 91.0\%. The corresponding VQE share is 90.9\%. Removing B from the full VQE configuration increases runtime by $4.19\times$, and removing P increases it by $2.59\times$. Reducing the cost of search and the amount of search are both important.

Adaptive arithmetic and output construction provide the largest saving for PQF. Along the same B--P--A sequence, adding A accounts for 76.4\% of the time saved on random workloads and 81.1\% of the time saved on VQE workloads. Removing A from the full VQE configuration makes synthesis $10.44\times$ slower. PQF constructs a primary circuit, a fallback circuit, and numerical fields for their probabilities and errors. Specialized arithmetic, shared constants, and faster decimal conversion reduce the cost of producing this complete result.

The benefit of each optimization depends on which other groups are enabled. On VQE PQF workloads, pruning alone takes 101.6\% of the original time, while removing it from the full configuration increases runtime by 7.8\%. Combining B and P reduces random RustGridSynth time to 18.5\% of the original, compared with 59.5\% for B alone and 82.0\% for P alone. Together, these results show that pruning contributes more when combined with the other optimizations.

\subsection{RQ3: Circuit Cost and Memory}

Atlas meets the requested error tolerance with aggregate mean $T$-cost changes below 0.31\% and sometimes even better. For VQE, RustGridSynth's mean changes by $+0.27\%$ on EPYC and $-0.19\%$ on Xeon; PQF's expected cost decreases by 1.23\% and 0.61\%, respectively. Across random workloads, RustGridSynth's changes remain within 0.51\%, and PQF's expected cost decreases by $0.86$--$0.87\%$. The benchmark harness accepts timing records after checking exact unitarity and approximation error for deterministic outputs, and branch errors, probabilities, and expected cost for PQF.

The additional memory overhead by Atlas is modest. Memory use increases most for PQF, whose mean summed worker RSS rises from 48.7 to 102.9\,MiB on EPYC's VQE workloads. The corresponding Xeon values are 47.6 and 73.5\,MiB. RustGridSynth changes more modestly: from 69.5 to 78.2\,MiB on EPYC and from 65.7 to 63.5\,MiB on Xeon. We sample resident set size approximately every 10\,ms in a separate run, sum it across workers, and average the per-case means.

\subsection{RQ4: Effect on Compilation Time}

We estimate compilation time by combining measured synthesis time with measured mapping and circuit-optimization time. We measure layout, routing, translation, and circuit optimization with Qiskit \cite{qiskit2024} (for general circuit optimizations) + LSC \cite{watkins2024high} (for FTQC routing), mapping the ten VQE circuits onto nearest-neighbor square grids: $4\times4$ for 10 logical qubits and $5\times5$ for 20. Each source Pauli rotation expands into basis changes, a CNOT ladder, and an RZ gate; mapping retains the H/RZ/CX basis.

Atlas reduces end-to-end compilation time by $1.97$--$2.37\times$ with RustGridSynth synthesis at $\varepsilon=10^{-8}$. Table~\ref{tab:eval-compilation} gives the totals for both CPUs and methods. On EPYC, mapping takes 223.9 seconds, while RustGridSynth synthesis falls from 297.0 to 40.7 seconds. Their sum therefore falls from 520.9 to 264.6 seconds. The PQF estimates improve by $6.94\times$ on EPYC and $4.33\times$ on Xeon, reflecting its larger synthesis component.

\begin{table}[t]
\centering
\caption{End-to-end compilation time for all ten VQE circuits at $\varepsilon=10^{-8}$. Times are in seconds.}
\label{tab:eval-compilation}
\begin{tabular}{lrrr}
\hline
CPU / method & Original & Atlas & Speedup\\
\hline
EPYC / RustGridSynth & 520.9 & 264.6 & $1.97\times$\\
Xeon / RustGridSynth & 305.0 & 128.5 & $2.37\times$\\
EPYC / PQF & 3,313.2 & 477.1 & $6.94\times$\\
Xeon / PQF & 2,377.4 & 548.9 & $4.33\times$\\
\hline
\end{tabular}
\end{table}

\begin{figure}[t]
\centering
\includegraphics[width=\columnwidth]{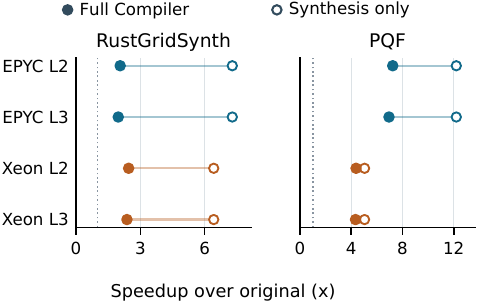}
\caption{Synthesis speedup and compilation speedup for ten VQE circuits at $\varepsilon=10^{-8}$. L2 and L3 denote Qiskit optimization levels 2 and 3.}
\label{fig:eval-compilation}
\end{figure}

Hardware mapping and circuit optimization dominate the remaining compilation estimate after rotation synthesis is accelerated. At the same tolerance and level, these passes account for 84.6\% of the Atlas compilation estimate on EPYC and 74.7\% on Xeon. Figure~\ref{fig:eval-compilation} shows the resulting gap between synthesis-only speedup and the end-to-end compilation pipeline at both optimization levels. Mapping and optimization are therefore the next targets for larger compilation gains.

Overall, for VQE applications, compilation results excluding rotation synthesis can be reused across the training process, making them one-time offline cost. Instead, rotation synthesis is the online cost across VQE training loop and deserves dedicated compilation accelerations.

%% file: sections/discussion.tex
\section{Discussion and Conclusion}\label{sec:discussion}

Quantum rotation synthesis may lead to remarkable overhead in quantum circuit compilation. 
Considering variational quantum eigensolver (VQE) applications as an example, a conventional UCCSD ansatz in quantum chemistry can generate $O(n^4)$ rotations for $n$ qubits. Each optimization iteration updates the rotation angles, making rotation synthesis a recurring cost that can delay the VQE training loop. We argue for the additional importance of accelerating the rotation synthesis process: for VQE, its circuit structure preserves and we can reuse compilation results such as qubit routing and gate optimizations; yet impossible for rotation gate synthesis where angles dramatically change.

We identify two sources of delay in rotation synthesis: redundant computation during synthesis and uneven synthesis times that leave workers waiting for slow requests. Atlas addresses these costs with three optimizations: search pruning skips invalid synthesis candidates through statistical inference; adaptive arithmetic avoid unnecessary high-overhead big integer/float arithmetic when decomposing the target rotation gate; batch scheduling fills the gap between the synthesis time of varied rotations. 
Across various benchmarks, synthesis method, and compilation machines, Atlas can consistently accelerate the rotation synthesis process by up to 12.65x, compared to the state-of-the-art.



%% file: main.bbl
\begin{thebibliography}{10}
\providecommand{\url}[1]{#1}
\csname url@samestyle\endcsname
\providecommand{\newblock}{\relax}
\providecommand{\bibinfo}[2]{#2}
\providecommand{\BIBentrySTDinterwordspacing}{\spaceskip=0pt\relax}
\providecommand{\BIBentryALTinterwordstretchfactor}{4}
\providecommand{\BIBentryALTinterwordspacing}{\spaceskip=\fontdimen2\font plus
\BIBentryALTinterwordstretchfactor\fontdimen3\font minus \fontdimen4\font\relax}
\providecommand{\BIBforeignlanguage}[2]{{%
\expandafter\ifx\csname l@#1\endcsname\relax
\typeout{** WARNING: IEEEtran.bst: No hyphenation pattern has been}%
\typeout{** loaded for the language `#1'. Using the pattern for}%
\typeout{** the default language instead.}%
\else
\language=\csname l@#1\endcsname
\fi
#2}}
\providecommand{\BIBdecl}{\relax}
\BIBdecl

\bibitem{ross2016optimal}
\BIBentryALTinterwordspacing
N.~J. Ross and P.~Selinger, ``Optimal ancilla-free {Clifford+T} approximation of {$z$}-rotations,'' \emph{Quantum Information and Computation}, vol.~16, no. 11--12, pp. 901--953, 2016. [Online]. Available: \url{https://arxiv.org/abs/1403.2975}
\BIBentrySTDinterwordspacing

\bibitem{bocharov2015pqf}
\BIBentryALTinterwordspacing
A.~Bocharov, M.~Roetteler, and K.~M. Svore, ``Efficient synthesis of probabilistic quantum circuits with fallback,'' \emph{Physical Review A}, vol.~91, no.~5, p. 052317, 2015. [Online]. Available: \url{https://doi.org/10.1103/PhysRevA.91.052317}
\BIBentrySTDinterwordspacing

\bibitem{peruzzo2014variational}
\BIBentryALTinterwordspacing
A.~Peruzzo, J.~McClean, P.~Shadbolt, M.-H. Yung, X.-Q. Zhou, P.~J. Love, A.~Aspuru-Guzik, and J.~L. O'Brien, ``A variational eigenvalue solver on a photonic quantum processor,'' \emph{Nature Communications}, vol.~5, p. 4213, 2014. [Online]. Available: \url{https://doi.org/10.1038/ncomms5213}
\BIBentrySTDinterwordspacing

\bibitem{mcclean2016theory}
\BIBentryALTinterwordspacing
J.~R. McClean, J.~Romero, R.~Babbush, and A.~Aspuru-Guzik, ``The theory of variational hybrid quantum-classical algorithms,'' \emph{New Journal of Physics}, vol.~18, no.~2, p. 023023, 2016. [Online]. Available: \url{https://doi.org/10.1088/1367-2630/18/2/023023}
\BIBentrySTDinterwordspacing

\bibitem{romero2019ucc}
\BIBentryALTinterwordspacing
J.~Romero, R.~Babbush, J.~R. McClean, C.~Hempel, P.~J. Love, and A.~Aspuru-Guzik, ``Strategies for quantum computing molecular energies using the unitary coupled cluster ansatz,'' \emph{Quantum Science and Technology}, vol.~4, no.~1, p. 014008, 2019. [Online]. Available: \url{https://doi.org/10.1088/2058-9565/aad3e4}
\BIBentrySTDinterwordspacing

\bibitem{ryabinkin2018qcc}
\BIBentryALTinterwordspacing
I.~G. Ryabinkin, T.-C. Yen, S.~N. Genin, and A.~F. Izmaylov, ``Qubit coupled cluster method: A systematic approach to quantum chemistry on a quantum computer,'' \emph{Journal of Chemical Theory and Computation}, vol.~14, no.~12, pp. 6317--6326, 2018. [Online]. Available: \url{https://doi.org/10.1021/acs.jctc.8b00932}
\BIBentrySTDinterwordspacing

\bibitem{maldacena2016syk}
\BIBentryALTinterwordspacing
J.~Maldacena and D.~Stanford, ``Remarks on the {Sachdev--Ye--Kitaev} model,'' \emph{Physical Review D}, vol.~94, no.~10, p. 106002, 2016. [Online]. Available: \url{https://doi.org/10.1103/PhysRevD.94.106002}
\BIBentrySTDinterwordspacing

\bibitem{kliuchnikov2013exact}
\BIBentryALTinterwordspacing
V.~Kliuchnikov, D.~Maslov, and M.~Mosca, ``Fast and efficient exact synthesis of single-qubit unitaries generated by {Clifford} and {T} gates,'' \emph{Quantum Information and Computation}, vol.~13, no. 7--8, pp. 607--630, 2013. [Online]. Available: \url{https://arxiv.org/abs/1206.5236}
\BIBentrySTDinterwordspacing

\bibitem{paetznick2014rus}
\BIBentryALTinterwordspacing
A.~Paetznick and K.~M. Svore, ``{Repeat-Until-Success}: Non-deterministic decomposition of single-qubit unitaries,'' \emph{Quantum Information and Computation}, vol.~14, no. 15--16, pp. 1277--1301, 2014. [Online]. Available: \url{https://www.rintonpress.com/xxqic14/qic-14-1516/1277-1301.pdf}
\BIBentrySTDinterwordspacing

\bibitem{campbell2017mixing}
\BIBentryALTinterwordspacing
E.~Campbell, ``Shorter gate sequences for quantum computing by mixing unitaries,'' \emph{Physical Review A}, vol.~95, no.~4, p. 042306, 2017. [Online]. Available: \url{https://doi.org/10.1103/PhysRevA.95.042306}
\BIBentrySTDinterwordspacing

\bibitem{kliuchnikov2023shorter}
\BIBentryALTinterwordspacing
V.~Kliuchnikov, K.~Lauter, R.~Minko, A.~Paetznick, and C.~Petit, ``Shorter quantum circuits via single-qubit gate approximation,'' \emph{Quantum}, vol.~7, p. 1208, 2023. [Online]. Available: \url{https://quantum-journal.org/papers/q-2023-12-18-1208/}
\BIBentrySTDinterwordspacing

\bibitem{morisaki2025optimal}
\BIBentryALTinterwordspacing
H.~Morisaki, K.~Sano, and S.~Akibue, ``Optimal ancilla-free {Clifford+T} synthesis for general single-qubit unitaries,'' 2025, arXiv preprint. [Online]. Available: \url{https://arxiv.org/abs/2510.05816}
\BIBentrySTDinterwordspacing

\bibitem{bothe2026smallangle}
\BIBentryALTinterwordspacing
M.~Bothe, C.~S{\"u}nderhauf, M.~J. Witham, E.~T. Campbell, and N.~S. Blunt, ``More efficient {Clifford+T} synthesis for small-angle rotations and application to {Trotterization},'' 2026, arXiv preprint, version 2. [Online]. Available: \url{https://arxiv.org/abs/2605.31544}
\BIBentrySTDinterwordspacing

\bibitem{nam2018optimization}
\BIBentryALTinterwordspacing
Y.~Nam, N.~J. Ross, Y.~Su, A.~M. Childs, and D.~Maslov, ``Automated optimization of large quantum circuits with continuous parameters,'' \emph{npj Quantum Information}, vol.~4, no.~1, p.~23, 2018. [Online]. Available: \url{https://doi.org/10.1038/s41534-018-0072-4}
\BIBentrySTDinterwordspacing

\bibitem{kishony2026optimal}
\BIBentryALTinterwordspacing
G.~Kishony, A.~Elazari, R.~Cohen, and L.~Gazit, ``Optimal approximation of single qubit rotations within a quantum circuit,'' 2026, arXiv preprint. [Online]. Available: \url{https://arxiv.org/abs/2605.09671}
\BIBentrySTDinterwordspacing

\bibitem{amy2013mitm}
\BIBentryALTinterwordspacing
M.~Amy, D.~Maslov, M.~Mosca, and M.~Roetteler, ``A meet-in-the-middle algorithm for fast synthesis of depth-optimal quantum circuits,'' \emph{IEEE Transactions on Computer-Aided Design of Integrated Circuits and Systems}, vol.~32, no.~6, pp. 818--830, 2013. [Online]. Available: \url{https://doi.org/10.1109/TCAD.2013.2244643}
\BIBentrySTDinterwordspacing

\bibitem{dimatteo2016parallelizing}
\BIBentryALTinterwordspacing
O.~Di~Matteo and M.~Mosca, ``Parallelizing quantum circuit synthesis,'' \emph{Quantum Science and Technology}, vol.~1, no.~1, p. 015003, 2016. [Online]. Available: \url{https://doi.org/10.1088/2058-9565/1/1/015003}
\BIBentrySTDinterwordspacing

\bibitem{paradis2024synthetiq}
\BIBentryALTinterwordspacing
A.~Paradis, J.~Dekoninck, B.~Bichsel, and M.~Vechev, ``{Synthetiq}: Fast and versatile quantum circuit synthesis,'' \emph{Proceedings of the ACM on Programming Languages}, vol.~8, no. OOPSLA1, pp. 96:1--96:28, 2024. [Online]. Available: \url{https://doi.org/10.1145/3649813}
\BIBentrySTDinterwordspacing

\bibitem{glaudell2026highperformance}
\BIBentryALTinterwordspacing
A.~N. Glaudell, M.~Jarret, S.~Klein, S.~S. Mendelson, T.~C. Mooney, and M.~Tian, ``High-performance exact synthesis of two-qubit quantum circuits,'' 2026, arXiv preprint, version 1. [Online]. Available: \url{https://arxiv.org/abs/2601.19166}
\BIBentrySTDinterwordspacing

\bibitem{cudaqRotationSynthesis}
\BIBentryALTinterwordspacing
{NVIDIA}, ``Rotation synthesis ({Clifford+T}),'' {CUDA-Q} documentation, n.d., accessed November 19, 2026. [Online]. Available: \url{https://nvidia.github.io/cuda-quantum/latest/using/examples/rotation_synthesis.html}
\BIBentrySTDinterwordspacing

\bibitem{montgomery1985modular}
P.~L. Montgomery, ``Modular multiplication without trial division,'' \emph{Mathematics of Computation}, vol.~44, no. 170, pp. 519--521, 1985.

\bibitem{brent1980improved}
R.~P. Brent, ``An improved {Monte Carlo} factorization algorithm,'' \emph{BIT Numerical Mathematics}, vol.~20, pp. 176--184, 1980.

\bibitem{rsgridsynth2026}
\BIBentryALTinterwordspacing
S.~Yamamoto, N.~Yoshioka, and {IBM}, ``{rsgridsynth},'' Rust software library, version 0.2.2, 2026, released August 1, 2026. Accessed November 19, 2026. [Online]. Available: \url{https://docs.rs/rsgridsynth/0.2.2/rsgridsynth/}
\BIBentrySTDinterwordspacing

\bibitem{qiskit2024}
\BIBentryALTinterwordspacing
A.~Javadi-Abhari, M.~Treinish, K.~Krsulich, C.~J. Wood, J.~Lishman, J.~Gacon, S.~Martiel, P.~D. Nation, L.~S. Bishop, A.~W. Cross, B.~R. Johnson, and J.~M. Gambetta, ``Quantum computing with {Q}iskit,'' 2024, arXiv:2405.08810. [Online]. Available: \url{https://arxiv.org/abs/2405.08810}
\BIBentrySTDinterwordspacing

\bibitem{watkins2024high}
\BIBentryALTinterwordspacing
G.~Watkins, H.~M. Nguyen, K.~Watkins, S.~Pearce, H.-K. Lau, and A.~Paler, ``A high performance compiler for very large scale surface code computations,'' \emph{Quantum}, vol.~8, p. 1354, 2024. [Online]. Available: \url{https://doi.org/10.22331/q-2024-05-22-1354}
\BIBentrySTDinterwordspacing

\end{thebibliography}
